\documentclass[prl,aps,twocolumn,groupedaddress]{revtex4}
\usepackage{epsfig,latexsym,amssymb,amsmath,amsbsy,graphics,graphicx}

\def \beq{\begin{equation}}
\def \eeq{\end{equation}}
\def \beqarr{\begin{eqnarray}}
\def \eeqarr{\end{eqnarray}}

\begin{document}

\title{Quantum criticality of a Fermi gas with a spherical dispersion minimum}

\author{Kun Yang}
\thanks{On leave from Department of Physics, Florida State University,
Tallahassee, FL 32306.}

\affiliation{Department of Physics, Harvard University, Cambridge,
MA 02138, USA}

\author{Subir Sachdev}

\affiliation{Department of Physics, Harvard University, Cambridge,
MA 02138, USA}

\date{November 27, 2005}

\begin{abstract}

We describe the quantum phase transition of a Fermi gas occurring
when the quasiparticle excitation energy has a minimum in momentum
space which crosses zero on a sphere of radius $k_0 \neq 0$. The
quasiparticles have a universal interaction which controls the
physical properties in vicinity of the quantum critical point. We
discuss possible applications to fermionic superfluids formed by
pairing two fermion species, near the point where the densities of
the two species become unequal.

\end{abstract}

\maketitle

The study of quantum and thermal fluctuations near second-order,
zero temperature quantum phase transitions (``quantum criticality'')
has been an important theme in modern condensed matter physics
\cite{coleman}, providing a unifying framework for many experimental
studies of heavy fermion compounds and correlated oxides. With the
advent of experiments on trapped, quantum degenerate, ultracold
atoms a new arena for the study of tunable quantum phase transitions
has opened up; the most prominent example being the study of
superfluid-insulator transition of bosonic $^{87}$Rb atoms in an
optical lattice\cite{greiner}. Experiments have also studied the
condensate formed by pairs of fermionic $^{6}$Li atoms with distinct
hyperfine states, and very recently these have been extended to a
quantum phase transition into a Fermi liquid by unbalancing the
density of the hyperfine states \cite{martin}.

This paper will further explore the physics of paired fermion
systems with unequal densities of the two fermion species. Apart
from cold atom systems, such problems are also of interest to
studies of electronic superconductors in an applied magnetic field
\cite{radovan}, and to the formation of paired quark condensates at
high nucleon densities \cite{review}. We will introduce and solve a
theoretical model of a quantum phase transition that occurs when the
densities of the fermion species is initially unbalanced, and
describe the universal physical properties in its vicinity. The
basis of our analysis will be a second-order quantum critical point
(QCP) which allows a systematic renormalization group analysis of
arbitrary fermion interactions in its vicinity. We will show that
the QCP is a powerful and unifying vantage point for describing a
variety of strongly interacting phases obtained as the fermion
densities and temperature are varied.

The essential characteristic of our model is that there is a
conserved U(1) ``charge'', $\mathcal{Q}$, which is carried by the
underlying fermions; it is also required that the paired fermion
condensate is neutral under this U(1) symmetry. For the ultracold
atom systems, $\mathcal{Q}$ is the difference in the density of the
two fermion species, while for electronic superconductors in an
applied magnetic field, $\mathcal{Q}$ is the total spin $S_Z$ along
the field direction. Now consider the situation at zero temperature
($T$) as a chemical potential $\mu$ conjugate to $\mathcal{Q}$ is
varied \cite{note}. We are interested in a QCP at which $\langle
\mathcal{Q}\rangle$ has a non-analytic dependence on $\mu$
\cite{book}; by a shift in $\mu$, we choose this QCP to be at
$\mu=0$. Typically, $\langle \mathcal{Q} \rangle$ is independent of
$\mu$ for $\mu < 0$, and increases smoothly with $\mu$ for $\mu >
0$.

\begin{figure}
\centering
\includegraphics[width=3.4in]{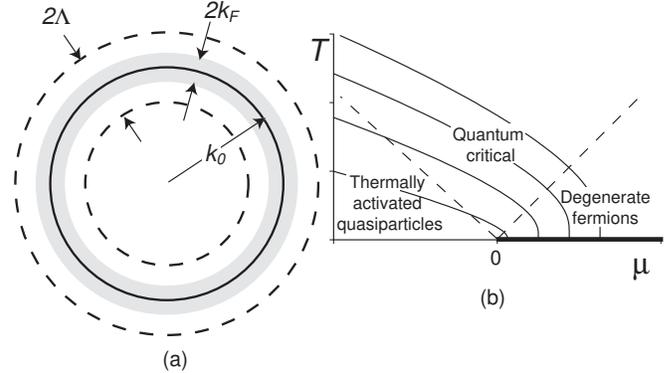}
\caption{Properties of $H$. ({\em a\/}) For $\mu < 0$ the ground
state of $H$ is the fermion vacuum, while for $\mu>0$ the ground
state has fermions occupying the shaded region with $k_0-k_F
<k<k_0+k_F$. $\Lambda$ is a high momentum cutoff. ({\em b\/})
Crossover phase diagram: the dashed lines indicate crossovers
between the labeled regimes, and the universal interaction applies
in all three regimes. The full lines are contours of constant
$\langle \mathcal{Q} \rangle$. In applications to paired fermion
problems, all phases are superfluid, and the FFLO or
``breached-pair'' phases appear at low $T$ in the degenerate fermion
regime. Note that for experiments at fixed $\langle \mathcal{Q}
\rangle$ (as in cold atoms) there is a substantial intermediate
range of $T$ in the quantum critical region, even though the ground
state is in the degenerate fermion regime. } \label{circles}
\end{figure}

We will describe the situation where for $\mu<0$ there is a gapped
fermionic quasiparticle excitation which has minimum in its
dispersion on a spherical shell of momentum ${\bf k}$ with $k = k_0
\neq 0$ (the $k_0=0$ case is considered in Ref.~\cite{book}). At
$\mu=0$, the minimum energy of this fermionic excitation vanishes,
and the onset in $\langle \mathcal{Q} \rangle$ variation is
accompanied by the appearance of two new Fermi surfaces at
wavevectors below and above $k_0$ (see Fig.~\ref{circles}a).
One of our primary results is that the interactions between low
lying fermion excitations at wavevectors near $k=k_0$ are universal,
and we will determine some of these interactions exactly.

It is worth noting here that solution of the conventional
weak-coupling Bardeen-Cooper-Schrieffer (BCS) theory does indeed
yield a minimum in the quasiparticle dispersion at a $k_0 \neq 0$.
However, application to the unbalanced fermion case does not yield a
QCP as described above: instead there is a strong first order
transition from a state with $\langle \mathcal{Q} \rangle =0$ to a
state with $\langle \mathcal{Q} \rangle \neq 0$ at some $\mu<0$
\cite{clogston}; similar first-order transitions have also been
obtained in recent mean-field analyses \cite{liu,son,sheehy}. On the
other hand, there is recent numerical evidence for a continuous
transition \cite{carlson}, in a situation with strong bare
interactions between the fermions and a renormalized quasiparticle
dispersion minimum at a $k_0 \neq 0$. We will find here that a
continuous transition can be present for generic short-range
interactions between the fermions. As is also the case for many
other second-order critical points \cite{balents}, long-wavelength
sound-mode fluctuations of the compressible environment lead to
marginally relevant flows and a likely weak fluctuation-induced
first order transition \cite{kun2}. Such a transition is best
described in the renormalization group framework introduced here,
and is distinct from that of the weak-coupling mean-field analyses.
In all cases, the underlying second-order QCP remains important for
understanding the full scope of the crossovers and transitions in
the $\mu$-$T$ plane.

We label the low-lying fermionic excitations in the vicinity of
$k=k_0$ by annihilation operators $c_{\bf k}$, so that $\mathcal{Q}
= \sum_{\bf k} c^\dagger_{\bf k} c_{\bf k}$; note that the fermions
do not carry any additional `flavor' or `spin' label associated with
quantum numbers distinct from $\mathcal{Q}$, but it is not difficult
to extend our analysis to include this. The universal quantum
critical properties can then be described by the following simple
Hamiltonian
\begin{equation}
H=\sum_{\bf k}[\epsilon(k)-\mu]c_{\bf k}^\dagger c_{\bf k} +{1\over
A} \sum_{{\bf k}, {\bf k}', {\bf q}}V_q c_{{\bf k}+ {\bf q}}^\dagger
c_{{\bf k}'-{\bf q}}^\dagger c_{{\bf k}'} c_{{\bf k}} ,
\label{model}
\end{equation}
where $V_q$ is a generic two-body interaction; $A$ is the volume of
the system, and the single-particle dispersion is
\begin{equation}
\epsilon(k)=(k-k_0)^2 / (2m^* ). \label{dispersion}
\end{equation}
For repulsive interactions, $H$ has a QCP at $\mu=0$, where the
fermion density increases continuously from zero as $\mu$ increases.
We will determine the critical behavior of the system at $\mu=0$,
and also study the properties of the system in the dilute limit
$k_F\approx\sqrt{2m^*|\mu|}\ll k_0$ (see Fig.~\ref{circles}a). In
two dimensions (2D) we find this low-density phase is a crystalline
phase with charge density wave (CDW) order; this corresponds to the
Fulde-Ferrell-Larkin-Ovchinnikov (FFLO) state\cite{ff,lo} in the
fermion pairing problem. On the other hand we find in 3D a uniform
phase is stable against CDW instability when the interaction is
sufficiently weak; this corresponds to the ``breached pairing''
phase \cite{liu}.

{\em The Quantum Critical Point} -- For simplicity we will focus
mainly on 2D, and briefly comment on the 3D case. Near the QCP at
$\mu=0$, the important low-energy fermionic modes are those lie
along a shell: $k_0 - \Lambda < k < k_0+\Lambda$, with the cutoff
$\Lambda\ll k_0$; these are the modes we keep in our model. Thus the
phase space structure and corresponding kinematic constraints of the
present problem is quite similar to that of the Fermi liquid fixed
point studied by Shankar \cite{shankar} using momentum shell
renormalization group (RG). There are, however, some important
differences. First, the quadratic fermion dispersion
(\ref{dispersion}) changes the scaling dimension of various
operators. Second, the ground state at the QCP is known exactly; it
is simply the fermion vacuum. This immediately leads to the exact
fermion $T=0$ Green's function (which is not modified by
interactions):
\begin{equation}
G(\omega, {\bf k})=1/[i\omega - \epsilon(k)]. \label{gg}
\end{equation}
The exact two fermion scattering vertices (or four-point functions)
$\Gamma({\bf k}_1, \omega_1, {\bf k}_1', \omega_1'; {\bf k}_1+{\bf
q}, \omega_1+\omega, {\bf k}_1'-{\bf q}, \omega_1'-\omega)$ can be
obtained simply by summing ladder diagrams. Here we need to divide
the scattering vertices into two classes. ({\em i\/}) The total
momentum of the two particles $|{\bf Q}|=|{\bf k}_1+{\bf k}_1'| \gg
\Lambda$; in this case kinematic constraints restrict the scattering
processes with momentum transfer $|{\bf q}|\lesssim \Lambda$, i.e.,
only forward scattering is possible in this case \cite{shankar}.
({\em ii\/}) $|{\bf Q}|=|{\bf k}_1+{\bf k}_1'| \ll \Lambda$; in this
case large angle scattering is possible, and these are the processes
in the Cooper channel \cite{shankar}.

For case ({\em i\/}) the one-loop correction to $\Gamma$ is
\begin{eqnarray}
 F^{(2)}(\theta,\Omega)=&-&\int{d^2{\bf q}\over (2\pi)^2}
{[V(\theta)]^2\over -i\Omega+{(|{\bf
k}_1+{\bf q}|-k_0)^2+(|{\bf k}_1'-{\bf q}|-k_0)^2\over 2m^*}}\nonumber\\
&=&-{m^*[V(\theta)]^2\over 2\pi|\sin\theta|}\log{\Lambda^2\over
2m^*|\Omega|},
\end{eqnarray}
where $\Omega=\omega_1+\omega_1'=\omega_2+\omega_2'$, $\theta$ is
the angle between ${\bf k}_1$ and ${\bf k}_1'$, and
$V(\theta)=V_{q=0}-V_{2k_0\sin(\theta/2)}$. The ladder sum is sum
over a geometric series that yields:
\begin{equation}
\Gamma=F(\theta,\Omega)=V(\theta)/\left[1+{m^*V(\theta)\over
2\pi|\sin\theta|}\log{\Lambda^2\over 2m^*|\Omega|}\right].
\label{Gammauniv}
\end{equation}
Clearly $F$ approaches a {\em universal\/} function in the
low-energy limit $\Omega\rightarrow 0$, in that it is independent of
the bare interaction $V$, when it is repulsive.

For case ({\em ii\/}), because ${\bf k}'\approx -{\bf k}$, it is
convenient to re-parameterize the vertex as $\Gamma({\bf k}_1,
\omega_1, -{\bf k}_1+{\bf Q}, \Omega-\omega_1; {\bf k}_2, \omega_2,
-{\bf k}_2+{\bf Q}, -\omega_2+\Omega)$, with $Q\ll \Lambda$, and
decompose it into angular momentum channels:
$\Gamma=\sum_n U_n(Q,\Omega)e^{in(\theta_{{\bf k}_1}-\theta_{{\bf
k}_2})}$.
A similar ladder sum yields
\begin{equation}
U_n(Q,\Omega)=V_n/\left\{1+V_nk_0\left[\sqrt{{m^*\over
4|\Omega|}}f\left({Q^2\over 2m^*|\Omega|}\right)-{m^*\over
\pi\Lambda}\right]\right\}, \label{un}
\end{equation}
 where $V_n={1\over 2\pi}\int{d\theta}e^{-in\theta}[V(\pi-\theta)-V(\theta)]$ (thus $n$ is odd), and
 $f(x)$ is a scaling function with the following asymptotic
 behavior: $f(x\rightarrow 0)\rightarrow 1$, and $f(x\rightarrow\infty)\rightarrow \sqrt{2\over \pi^2
 x}\log(x)$.
Again the vertex function takes a universal form in the limit
$\Omega, Q\rightarrow 0$ for repulsive interactions.

It is interesting to interpret the above results in the RG langauge.
The exact result in Eq.~(\ref{gg}) indicates that the dynamic
exponent $z=2$, the correlation length exponent $\nu=1/2$, and no
fermion anomalous dimension $\eta=0$. However, the presence of the
dimensionful scale $k_0$ means that, strictly speaking, there is no
true scale-invariant fixed point, and scaling arguments are of
limited utility. Nevertheless, it is possible to write down RG
equations and assign scaling dimensions for observables. The powers
of $k_0$ associated with an observable cannot be predicted {\em a
priori\/}, and explicit calculation is required. For forward
scattering we obtain the RG equation
\begin{equation}
{dV(\theta)\over d\log s}=-{m^*\over \pi |\sin\theta|}[V(\theta)]^2,
\label{forwardRG}
\end{equation}
where $s$ is the momentum re-scaling factor for $\Lambda$. This
implies forward scattering is marginally irrelevant/relevant for
repulsive/attractive interactions. The solution of
Eq.~(\ref{forwardRG}) also yields the universal structure in
Eq.~(\ref{Gammauniv}). In the Cooper channel, we have
\begin{equation}
{dV_n\over d\log s}=-{m^*k_0\over \pi\Lambda(s)}V_n^2.
\end{equation}
We can remove the explicit dependence of the $\beta$ function on
$\Lambda(s)$ by defining $\tilde{V}_n=V_n/\Lambda(s)$, and the flow
equation for $\tilde{V}_n$ then takes the form
\begin{equation}
{d\tilde{V}_n\over d\log s}=\tilde{V}_n-{m^*k_0\over
\pi}\tilde{V}_n^2.
\end{equation}
We thus find $\tilde{V}_n$ flows to a fixed point value of
${\pi\over m^*k_0}$ if it is repulsive initially. The physical $V_n$
is then obtained by rescaling with the observed frequency scale, a
conclusion consistent with Eq.~(\ref{un}).

The universal scattering vertices lead to a universal $T$ dependence
of the quasiparticle scattering rate (or inverse lifetime),
$1/\tau$, in the quantum-critical regime of Fig.~\ref{circles}b. The
large density of low energy states associated with the quadratic
quasiparticle dispersion leads to a short lifetime, and its value
has to be determined by computing the scattering rate into
quasiparticle states which have been self-consistently broadened. A
standard computation of the scattering cross-section between a
quasiparticle on the Fermi surface and pre-existing thermally
excited quasiparticle to order $\Gamma^2$ leads to the estimate
\begin{equation}
\frac{1}{\tau_{2D}} \sim T (m^\ast \Gamma)^2 (k_0^2/m^\ast)^{1/2}
[\mbox{Max}(T, 1/\tau_{2D})]^{-1/2}.
\end{equation}
The only self-consistent solution of this is
\begin{equation}
{1\over \tau_{2D}}={C_2 T^{2/3} (k_0^2/m^\ast )^{1/3} \over
|\log{\Lambda^2\over 2m^*T}|^{4/3}},
\end{equation}
where $C_2$ is a universal constant of order unity. The scattering
rate is universal in the sense that it is {\em independent} of the
bare interaction.

If the interaction is attractive, $\Gamma$ diverges at low energy in
all channels, and the strongest divergence is in the Cooper channel,
signaling the formation of bound states {\em before} the QCP at
$\mu=0$ is reached; such bound states appear at
$\mu=-V_\ell^2k_0^2m^*/8$, where $\ell$ is the angular momentum
channel with the strongest pairing interaction (i.e., $V_l$ being
the most negative).

Many of these results can be generalized to the 3D case with very
minor modifications. The most important difference between 2D and 3D
is in case (i), the forward scattering channel. While in 2D the
kinematic constraint only allows for small momentum transfer (or
forward scattering), in 3D the scattering process allows for a {\em
rotation} of the two momenta along the direction of ${\bf Q}$ while
preserving the angle between the two momenta \cite{shankar}, {\em
i.e.\/} ${\bf k}_1\cdot{\bf k}_1'\approx {\bf k}_2\cdot{\bf k}_1'$
or $\theta_{{\bf k}_1,{\bf k}_1'}\approx \theta_{{\bf k}_2,{\bf
k}_2'}\approx \theta$. The angle of rotation $\phi=\phi_{{\bf
k}_1}-\phi_{{\bf k}_2}$ ranges from 0 to $2\pi$. Thus in 3D the
scattering vertex is parameterized as $F(\theta, \phi, \Omega)$.
Performing a Fourier transform with respect to $\phi$:
$F(\theta, \phi, \Omega)=\sum_n F_n(\theta, \Omega)e^{in\phi}$,
and performing the ladder sum yields
\begin{equation}
F_n(\theta, \Omega)=V_n(\theta)/\left[1+{m^*k_0V_n(\theta)\over
4\pi|\cos(\theta/2)|}\log{\Lambda^2\over 2m^*|\Omega|}\right],
\end{equation}
where
\begin{equation}
V_n(\theta)={1\over 2\pi}\int_0^{2\pi}{d\phi}(V_{q(\theta,
\phi)}-V_{q(\theta, \pi-\phi)})e^{-in\phi}
\end{equation}
with $q(\theta, \phi)=2k_0\sin(\theta/2)\sin(\phi/2)$. While
$F_n(\theta, \Omega)$ approaches a universal function that does not
depend on the bare interaction in the low energy limit, generically
this is {\em not} the case for $F(\theta, \phi, \Omega)$. This is
because $F(\theta, \phi, \Omega)$ receives contributions from all
channels, and the energy scale below which $F_n(\theta, \Omega)$
approaches the universal function depends on $n$; this energy scale
approaches zero as $n\rightarrow \infty$. As a consequence the
quasiparticle scattering is nonuniversal in 3D; a straightforward
calculation similar to the 2D case finds that it is linear in $T$
with logarithmic corrections whose form depends on the details of
the bare interaction.

{\em The Low-density Phase} -- We now consider the phase with a
small $\mu$ or fermion density $\rho$. In the absence of
interaction, fermions occupy a thin ``Fermi shell" with $k_0-k_F < k
< k_0+k_F$, where the ``Fermi wavevector" $k_F= \pi\rho/k_0$ in 2D
and $\pi^2\rho/k_0^2$ in 3D is the half thickness of the shell (see
Fig~\ref{circles}a). In the following we show that this Fermi shell
state has a strong tendency toward charge density wave (CDW)
ordering, as reflected by enhanced static density susceptibilities
at certain wavevectors; the enhancement is due to the fact that the
``Fermi velocity" $v_F=k_F/m^*$ vanishes as $k_F\rightarrow 0$, but
the size of the Fermi surface remains finite and of order
$k_0^{d-1}$. We will show that in 2D this enhancement is strong
enough to render the system unstable against crystallization in the
low-density limit, in the presence of a {\em weak} repulsive
interactions between the fermions.

The strongest enhancement of static density susceptibility for
non-interacting fermions, $\chi^0(Q)$, is for $Q\approx 2k_F$, a
nesting vector that connects the inner and outer Fermi surfaces that
enclose the Fermi shell. A straightforward calculation for $Q\ll
k_0$ yields
\begin{equation}
\chi^0_{2D}(Q)={2m^*k_0\over k_F}K_{2D}\left({Q\over 2k_F}\right),
\end{equation}
where the dimensionless scaling function
\begin{equation}
K_{2D}(x)={1\over 16\pi^2x}\int_0^{2\pi}{d\theta\over
\cos\theta}\log\left|{1+x\cos\theta\over 1-x\cos\theta}\right|.
\end{equation}
The result is similar in 3D:
\begin{eqnarray}
\chi^0_{3D}(Q)&=&{2m^*k_0^2\over k_F}K_{3D}\left({Q\over
2k_F}\right),\\
K_{3D}(x)&=&{1\over 8\pi^2x}\int_0^1{dt}\log\left|{1+xt\over
1-xt}\right|.
\end{eqnarray}
Both $K_{2D}(x)$ and $K_{3D}(x)$ are sharply peaked at $x=1$. The
prefactor ${k_0/k_F}$ strongly enhances $\chi_0$ near the QCP; this
enhancement arises from the nearly perfect nesting between Fermi
surfaces which are curved in the same direction.

The other nesting wave vectors are $Q_{\pm}=2k_0\pm 2k_F$, which
connect the opposite ends of the outer and inner Fermi surfaces
respectively. We find
\begin{equation}
\chi^0_{2D}(Q_{\pm})={2m^*\over (2\pi)^2}\sqrt{k_0\over
k_F}\int_0^\infty{dt}{\log(1+{4\over t^2})\over 2+t^2}
\end{equation}
and $\chi^0_{3D}(Q_{\pm})=m^*k_0 /32$. We find $\chi^0(Q_{\pm})$
diverges in the low density limit in 2D while saturates in 3D.

Now let us go beyond non-interacting fermions, and include the
interactions computed earlier in a RPA-like theory of the
susceptibility $\chi$. Naively one might expect repulsive
interactions will immediately lead to a CDW instability at the
wavevector $Q=2k_F$ due to the divergent $\chi^0$ as $k_F\rightarrow
0$. However, there is a cancelation of interaction vertices in this
limit for such small momentum transfer; as a consequence the
strongest CDW instability is at {\em larger} wavevectors $Q_{\pm}$,
even though $\chi^0$ is less singular there. If we assume the $q$
dependence of $V_q$ is smooth over the scale of $2k_F$, which is the
case for short-range interactions, we can easily sum over the
particle-hole bubble and ladder diagrams and obtain
\begin{equation}
\chi(Q)\approx {\chi^0(Q)\over 1-(V_{q=0}-V_Q)\chi^0(Q)}.
\end{equation}
While $\chi^0(Q)$ is largest at $Q=2k_F$, interaction does not
immediately lead to an instability at this wavevector in the low
density limit, as for generic short-range interactions we expect
$V_{q=0}-V_Q\propto Q^2$ for small $Q$. We find instead that the
strongest density instability is at $Q=Q_{\pm}$, where the
interaction effects are stronger. In particular, as long as
$V_{q=0}-V_{2k_0} > 0$, in 2D $\chi_{2D}(Q_{\pm})$ {\em always}
diverges in the low-density limit $k_F\rightarrow 0$, suggesting
that the fermions form a charge density wave or crystal state in the
low density limit. Physically the CDW instability at wavevectors
that are {\em almost independent} of density in the low-density
limit can be understood in the following manner. In our model the
low-energy single fermion states are those with $k\approx k_0$; thus
localized states constructed from them must be oscillatory with a
characteristic wavevector $~ 2k_0$, similar to the Friedel
oscillation; this leads to strong CDW instability at such
wavevectors. We further note that in the low-density limit the
instability has equal strength for $Q_{\pm}=2k_0\pm 2k_F$; as a
consequence density modulations with characteristic wavevectors
$O(k_F)$ will be present due to interference. We expect their
magnitudes to be very large as $\chi$ is very strongly enhanced
there for small $k_F/k_0$. On the other hand the in 3D uniform state
appears to be stable against {\em weak interaction}, due to the fact
that $\chi^0_{3D}(Q_{\pm})$ is finite.

To summarize, we have presented a new perspective on the physics of
paired fermion systems, for the case in which the fermion densities
are unbalanced. It has recently become possible to study such
ultracold atom systems \cite{martin}, while condensed matter systems
have been under investigation for some time \cite{radovan,review}.
Essentially all previous analyses have been carried out in the
framework of extensions to BCS theory, seeking to minimize the free
energy with respect to variations in the magnitude and spatial
dependence of the order parameter. We have argued here that a more
rigorous approach is to focus on the interactions between the
renormalized quasiparticles as the system approaches a transition to
a gapless phase. This allowed us to make a number of controlled
predictions on the influence of strong interactions on the
quasiparticle dynamics and the structure of proximate phases in the
$\mu$-$T$ phase diagram.

This work was supported by NSF grants DMR-0225698 (KY) and
DMR-0537077 (SS).

\end{document}